\documentclass[11pt]{article}
\usepackage{amsmath}
\usepackage{amssymb}
\usepackage{amsthm}

\usepackage[T1]{fontenc}
\usepackage{textcomp}
\usepackage{yfonts}
\usepackage{graphicx}
\usepackage{color}
\usepackage{charter}
\usepackage{subfigure}

\definecolor{RedLetter}{rgb}{0.63,0.165,0.163}

\usepackage{amsmath}
\usepackage{amsbsy}
\usepackage{amssymb}
\usepackage{amsthm}
\usepackage[all,2cell,ps]{xy}

\newcommand{\vect}[1]{\boldsymbol{#1}}

\newcommand{\bF}{\vect{F}}
\newcommand{\bx}{\vect{x}}
\newcommand{\by}{\vect{y}}

\newcommand{\bdelta}{\vect{\delta}}
\newcommand{\balpha}{\vect{\alpha}}
\newcommand{\bbeta}{\vect{\beta}}

\newcommand{\refsec}[1]{Section \ref{#1}}
\newcommand{\refeqn}[1]{Eqn (\ref{#1})}
\newcommand{\reffig}[1]{Figure \ref{#1}}
\newcommand{\refthm}[1]{Theorem \ref{#1}}
\newcommand{\reflem}[1]{Lemma \ref{#1}}

\DeclareMathSymbol{\N}{\mathbin}{AMSb}{"4E}
\DeclareMathSymbol{\Z}{\mathbin}{AMSb}{"5A}
\DeclareMathSymbol{\R}{\mathbin}{AMSb}{"52}
\DeclareMathSymbol{\Q}{\mathbin}{AMSb}{"51}
\DeclareMathSymbol{\I}{\mathbin}{AMSb}{"49}

\newtheorem{thm}{Theorem}[section]
\newtheorem{lem}[thm]{Lemma}

\newcommand{\biblink}[2]{}

\newcommand{\ignore}[1]{}

\date{\today}

\newcommand{\ifeurovis}[1]{}

\begin{document}


\title{Dynamic Multilevel Graph Visualization}
\author{Todd L. Veldhuizen\footnote{Dept. of Electrical and Computer Engineering, University of Waterloo, Canada.  Email: {\tt tveldhui@acm.org}}}

\maketitle

\begin{abstract}
We adapt multilevel, force-directed graph layout techniques to
visualizing dynamic graphs in which vertices and edges are added
and removed in an online fashion (i.e., unpredictably).  
We maintain multiple levels of coarseness using
a dynamic, randomized coarsening algorithm.  To ensure the vertices
follow smooth trajectories, we employ dynamics simulation techniques,
treating the vertices as point particles.  We simulate fine and coarse
levels of the graph simultaneously, coupling the dynamics of
adjacent levels.  Projection from coarser to finer levels is adaptive,
with the projection determined by an affine transformation that
evolves alongside the graph layouts.  The result is a
dynamic graph visualizer that quickly and smoothly adapts to changes
in a graph.
\ifeurovis{
\begin{classification} 
\CCScat{H.5.0}{Information Systems}{Information Interfaces and Presentation: General}
\CCScat{G.2.2}{Discrete Mathematics}{Graph Theory: Graph Algorithms}
\end{classification}
}
\end{abstract}



\section{Introduction}


Our work is motivated by a
need to visualize dynamic graphs, that is, graphs from which vertices
and edges are being added and removed.
Applications include visualizing complex algorithms 
(our initial motivation), ad hoc wireless networks,
databases, monitoring distributed systems,
realtime performance profiling, and so forth.
Our design concerns are:
\begin{enumerate}
\item[D1.] The system should support \emph{online}
revision of the graph, that is, changes to the graph
that are not known in advance.  Changes made to the graph
may radically alter its structure.
\item[D2.] The animation should appear smooth.
It should be possible to visually track vertices
as they move, avoiding abrupt changes.
\item[D3.] Changes made to the graph should appear immediately,
and the layout should stabilize rapidly after a change.
\item[D4.] The system should produce aesthetically pleasing,
good quality layouts.
\end{enumerate}
We make two principle contributions:
\begin{enumerate}
\item We adapt multilevel force-directed graph layout algorithms
\cite{Walshaw:JGAA:2003} to the problem of
dynamic graph layout.
\item We develop and analyze an efficient algorithm for dynamically
maintaining the coarser versions of a graph needed for multilevel
layout.
\end{enumerate}

\subsection{Force-directed graph layout}

Forced-directed layout uses a physics metaphor to find 
graph layouts 
\cite{Eades:CN:1984,Kamada:IPL:1989,Frick:GD:1994,Fruchterman:SPE:1991}.
Each vertex is treated as a point particle
in a space (usually $\R^2$ or $\R^3$).  There are many
variations on how to translate the graph into physics.
We make fairly conventional choices, modelling
edges as springs which pull connected vertices 
together.  Repulsive forces between all pairs of
vertices act to keep the vertices spread out.  

We use a potential energy $V$ defined by\footnote{Somewhat
confusingly, we use $V$ for potential energy as well as
the set of vertices.  This is for consistency with 
Lagrangian dynamics.}
\begin{align}
V &= \underbrace{\sum_{(v_i,v_j) \in E} \tfrac{1}{2} K \|\bx_i - \bx_j\|^2}_{\text{spring potential}}
+ \underbrace{\sum_{v_i,v_j \in V, v_i \neq v_j } \tfrac{f_0}{\epsilon_R + \|\bx_i - \bx_j\|}}_{\text{repulsion potential}}
\label{e:potential}
\end{align}
where $\bx_i$ is the position of vertex $v_i$,
$K$ is a spring constant, $f_0$ is a
repulsion force constant, and $\epsilon_R$ is
a small constant used to avoid singularities.

To minimize the energy of \refeqn{e:potential},
one typically uses `trust region' methods, where
the layout is advanced in the general direction of the
gradient $\nabla V$, but restricting the distance by
which vertices may move in each step.  The maximum
move distance is often governed by an adaptive
`temperature' parameter
as in annealing methods, so that step sizes decrease
as the iteration converges.


One challenge in force-directed layout 
is that the repulsive forces that act to evenly
space the vertices become weaker as the graph becomes
larger.  This results in large graph layouts converging slowly, a
problem addressed by multilevel methods.

Multilevel graph layout algorithms
\cite{Walshaw:JGAA:2003,Koren:INFOVIS:2002} operate
by repeatedly `coarsening' a large graph to obtain a sequence of
graphs $G_0,G_1,\ldots,G_m$, where each $G_{i+1}$ has fewer
vertices and edges than $G_i$,
but is structurally similar.
For a pair $(G_i, G_{i+1})$, we refer to $G_i$ as the finer
graph and $G_{i+1}$ as the coarser graph.
The coarsest graph $G_m$ is laid
out using standard force-directed layout.  This layout is interpolated
(projected) to produce an initial layout for the finer graph $G_{m-1}$.
Once the force-directed layout of $G_{m-1}$ converges, it is
interpolated to provide an initial layout for $G_{m-2}$,
and so forth.

\subsection{Our approach}

Roughly speaking, we develop a dynamic 
version of Walshaw's multilevel force-directed layout
algorithm \cite{Walshaw:JGAA:2003}.

Because of criterion D3, that changes to the graph appear
immediately, we focused on approaches in which the
optimization process is visualized directly,
i.e., the vertex positions rendered reflect the
current state of the energy minimization process.

A disadvantage of the gradient-following algorithms
described above is that the layout can repeatedly overshoot a
minima of the potential function, resulting in zig-zagging.  
This is unimportant for offline layouts, 
but can result in jerky trajectories if the layout process is being animated.
We instead chose a dynamics-based approach in which vertices have
momentum.  Damping is used to
minimize oscillations.  This ensures that vertices follow
smooth trajectories (criterion D2).

We use standard dynamics techniques to simultaneously
simulate all levels of coarseness of a graph as one large 
dynamical system.
We couple the dynamics of each graph $V_i$ to its coarser version
$V_{i+1}$ so that `advice' about layouts can propagate 
from coarser to finer graphs.

Our approach entailed two major technical challenges:
\begin{enumerate}
\item How to maintain coarser versions of the graph
as vertices and edges are added and removed.
\item How to couple the dynamics of finer and coarser
graphs so that `layout advice' can quickly propagate from
coarser to finer graphs.
\end{enumerate}

We have addressed the first challenge by developing
a fully dynamic, Las Vegas-style randomized algorithm 
that requires $O(1)$ operations
per edge insertion or removal to maintain
a coarser version of a bounded degree
graph (\refsec{s:coarsening}).

We address the second challenge by using coarse graph vertices as
inertial reference frames for vertices in the fine graph
(\refsec{s:dynamics}).  The projection from coarser to finer graphs
is given dynamics, and evolves simultaneously with the vertex
positions, converging to a least-squares fit of the coarse graph
onto the finer graph (\refsec{s:framedynamics}).  We introduce time
dilations between coarser and finer graphs, which reduces the problem
of the finer graph reacting to cancel motions of the coarser graph
(\refsec{s:timedilation}).

\subsection{Demonstrations}

Accompanying this paper are the following movies.\footnote{
They may be downloaded from
\url{http://ubiety.uwaterloo.ca/ubigraph/ev/},
or by following the hyperlinks from
\reffig{f:demos}.
}
All movies are real-time screen captures on
an 8-core Mac.  Unless otherwise noted, the movies use one core
and 4th order Runge-Kutta integration.

\newcommand{\demo}[2]{\href{http://ubiety.uwaterloo.ca/ubigraph/ev/#1}{#2}}
\newcommand{\demoframe}[1]{\subfigure[ev1049\_{#1}]{\demo{ev1049_#1.mov}{\includegraphics[width=2.4in]{ev1049_#1.jpg}}}}
\begin{figure}
\begin{centering}
\demoframe{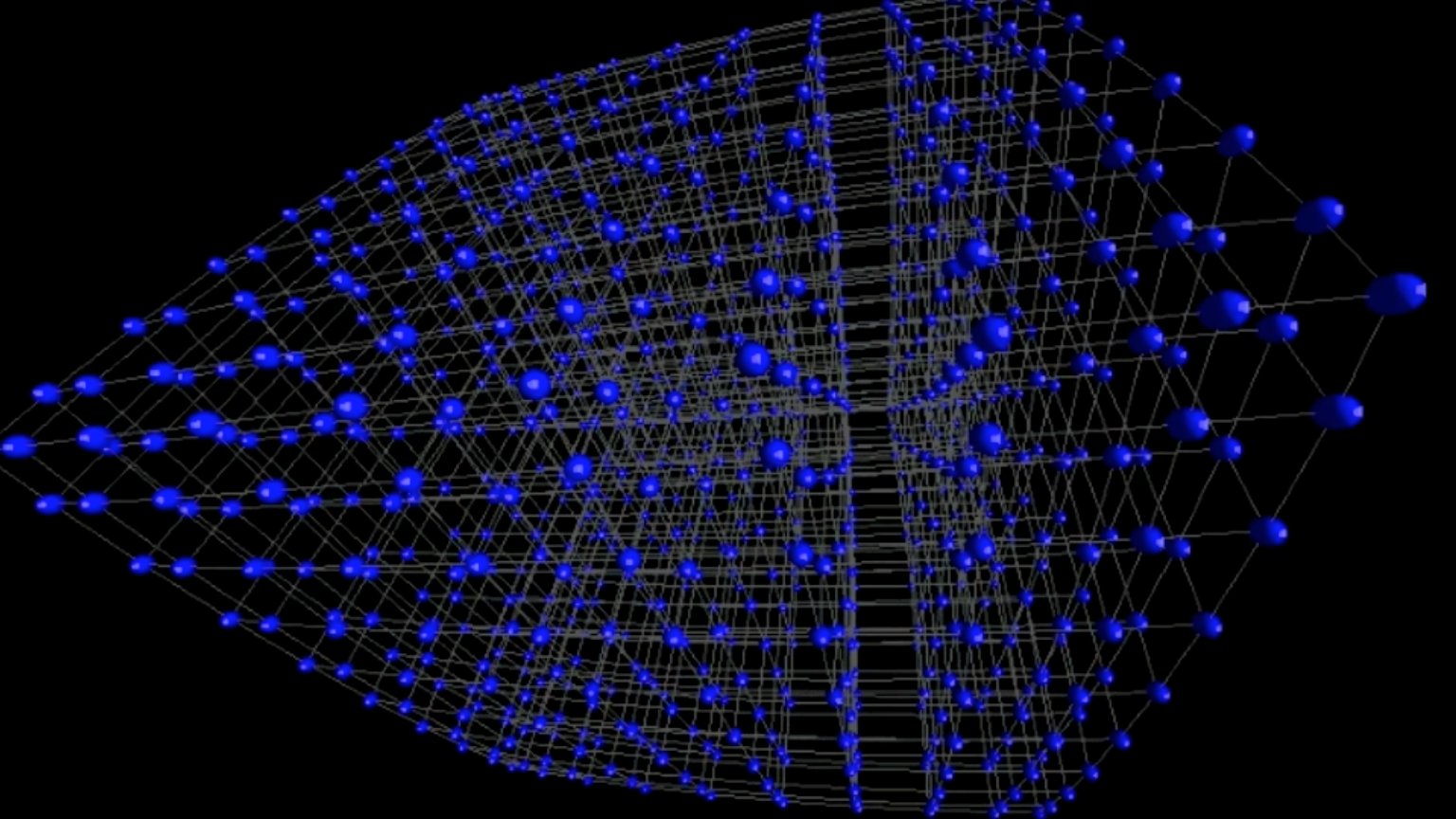}
\hspace{0.1in}
\demoframe{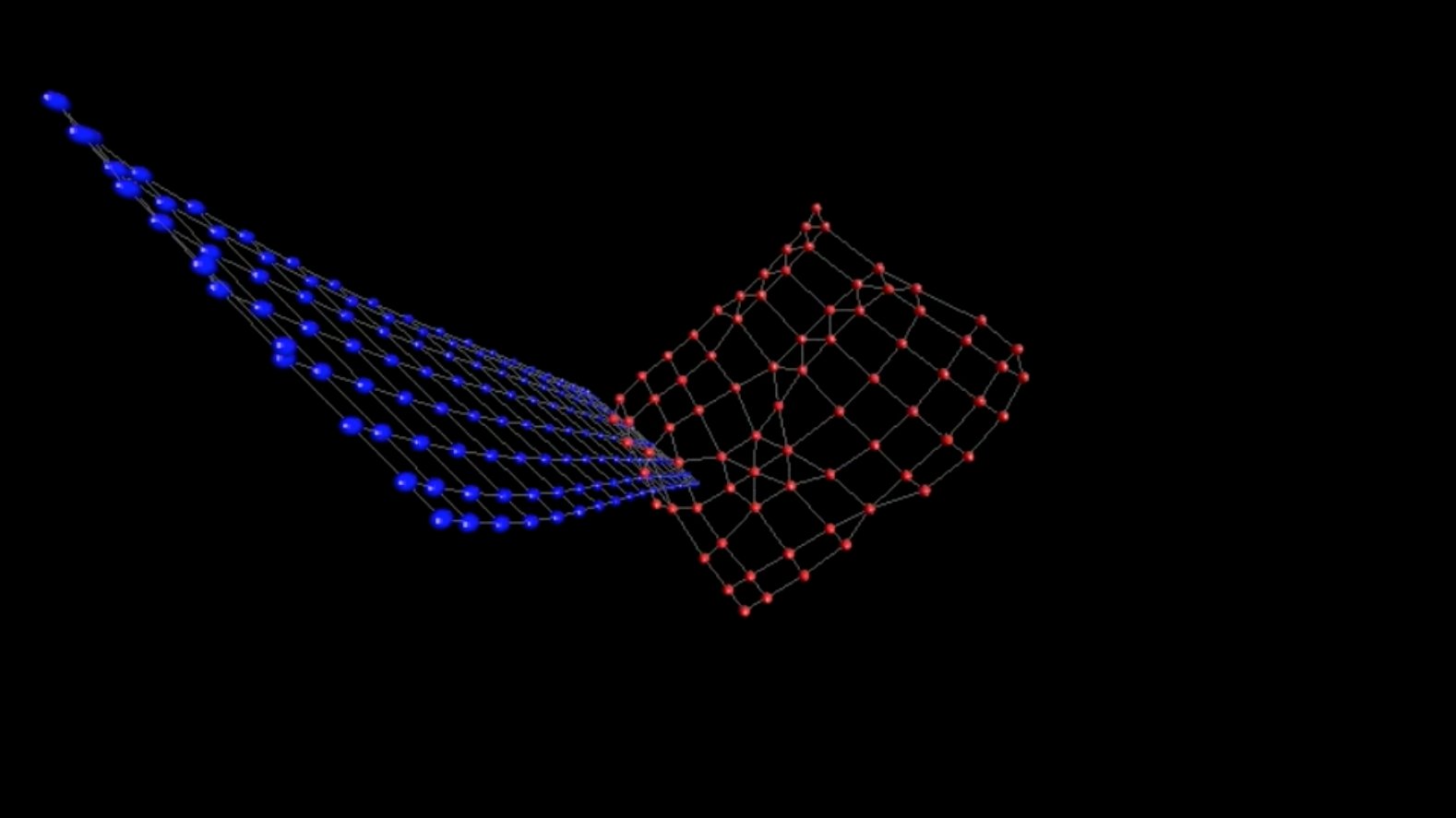}
\\
\demoframe{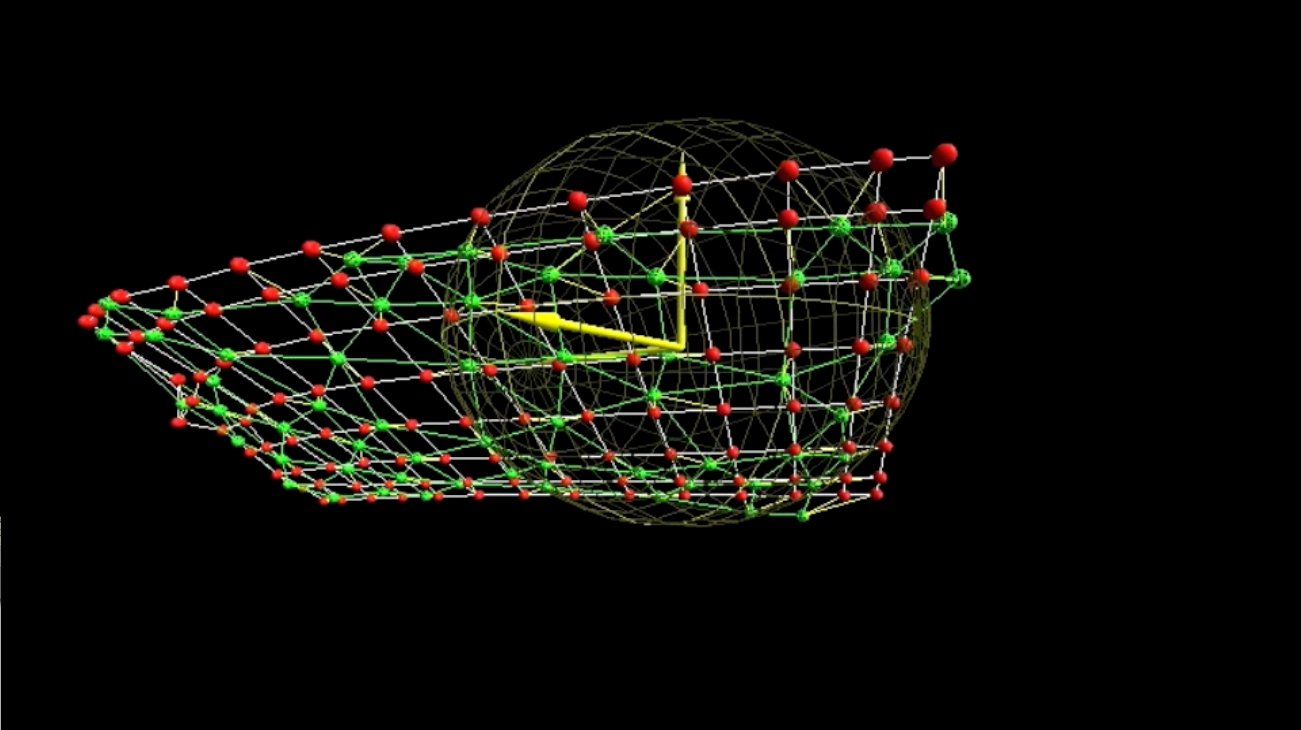}
\hspace{0.1in}\demoframe{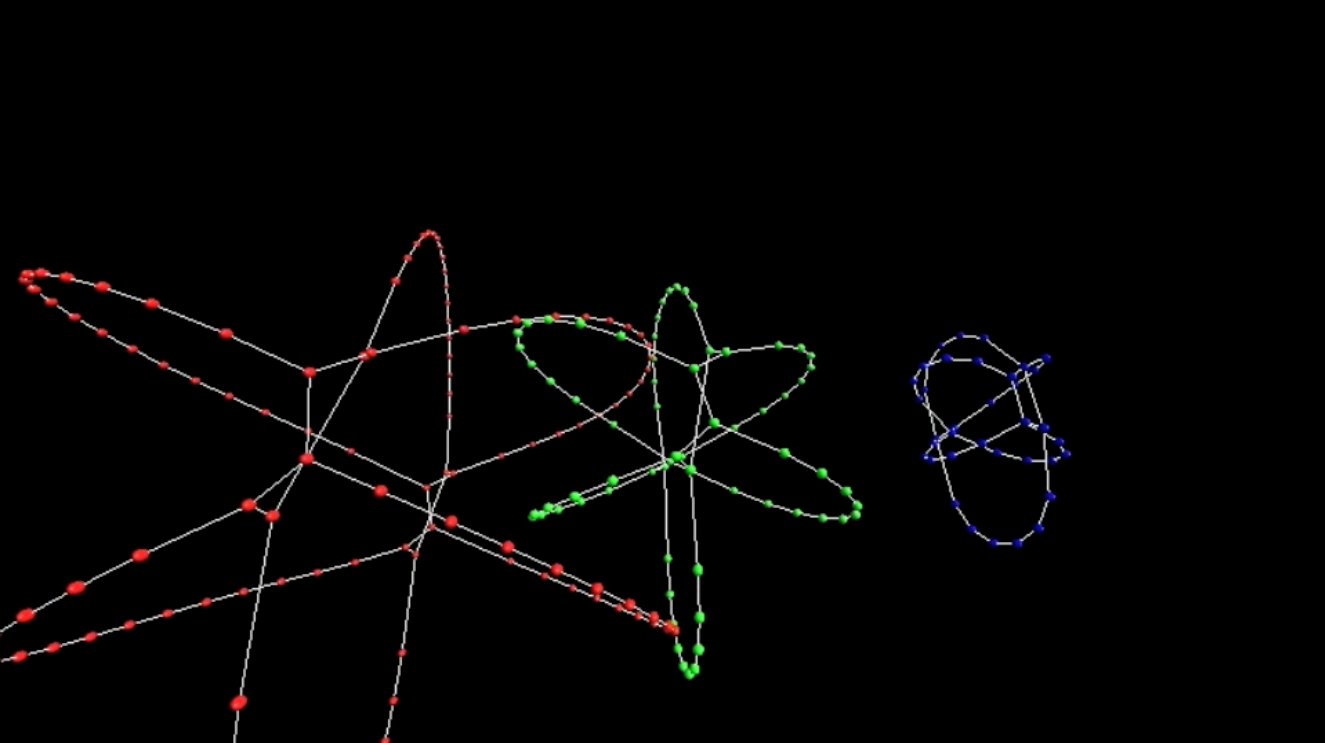}
\\
\demoframe{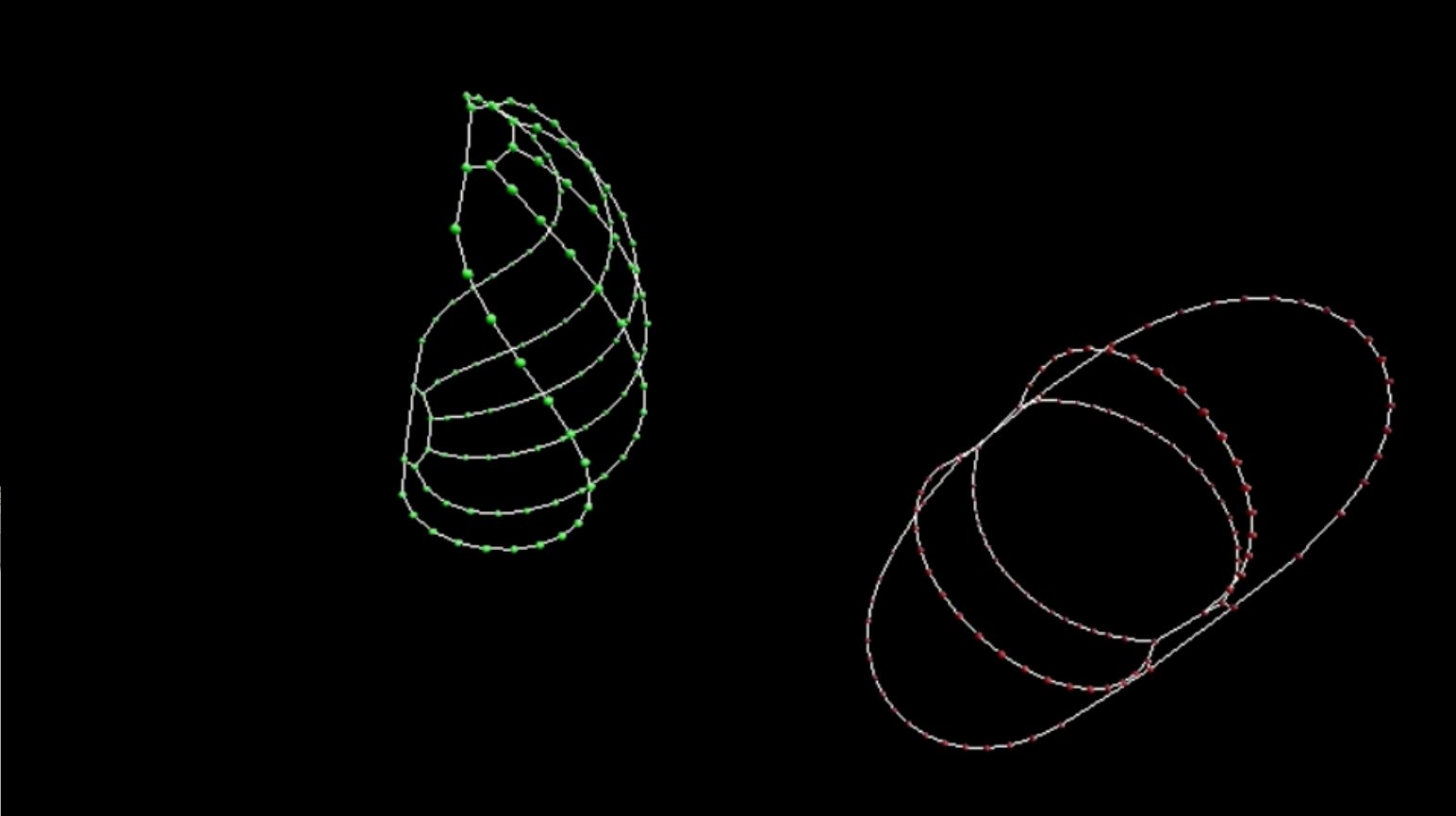}
\hspace{0.1in}
\demoframe{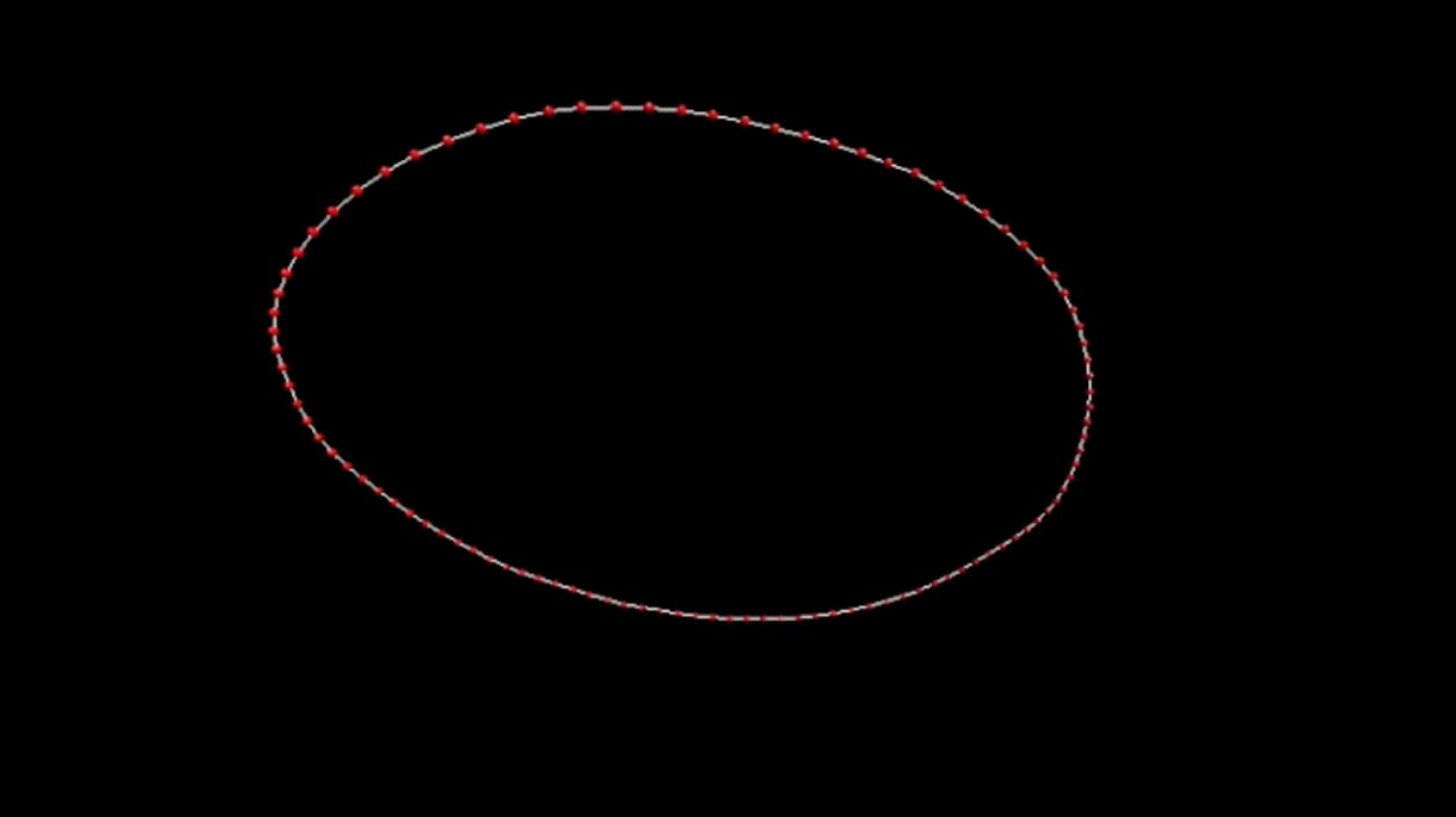}
\\
\demoframe{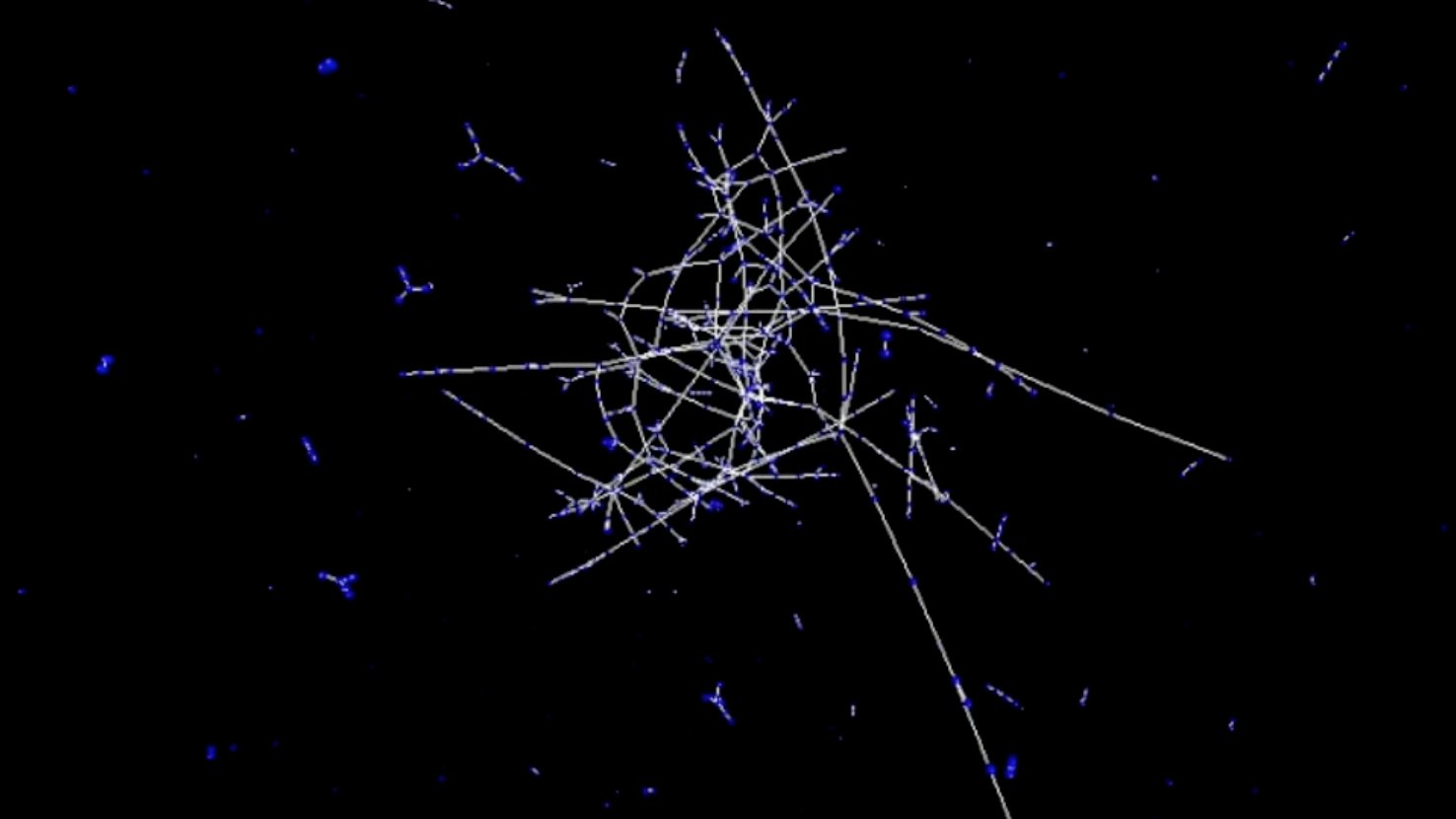}
\hspace{0.1in}\demoframe{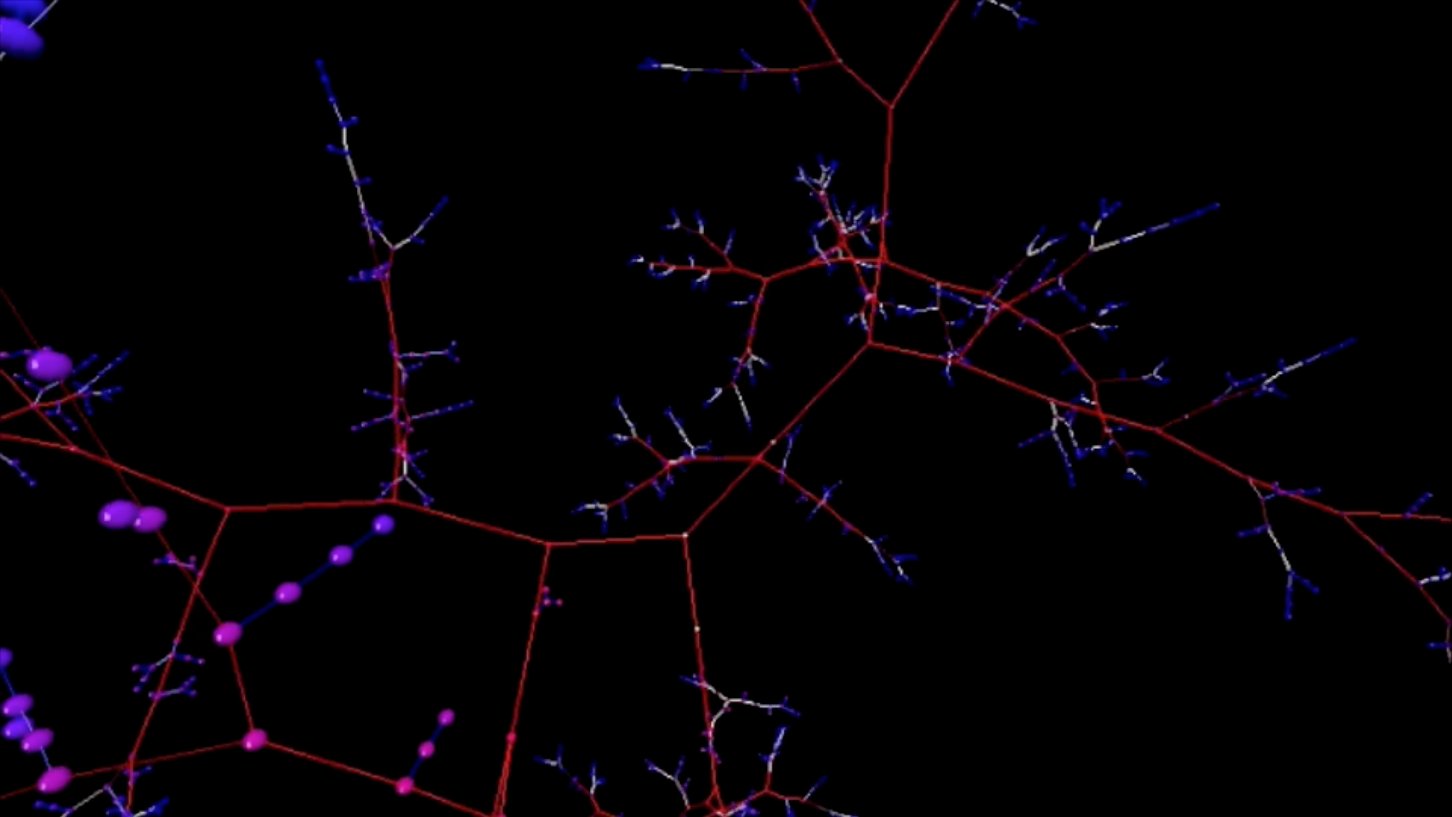}

\caption{\label{f:demos}Still frames from the demonstration movies accompanying this paper.}
\end{centering}
\end{figure}

\begin{itemize}
\item \demo{ev1049_cube.mov}{ev1049\_cube.mov}: Layout of a 10x10x10 cube graph using
single-level dynamics (\refsec{s:singleleveldynamics}), 8 cores, and 
Euler time steps.
\item \demo{ev1049_coarsening.mov}{ev1049\_coarsening.mov}: Demonstration of the dynamic
coarsening algorithm (\refsec{s:coarsening}).
\item \demo{ev1049_twolevel.mov}{ev1049\_twolevel.mov}: Two-level dynamics, showing
the projection dynamics and dynamic coarsening.
\item \demo{ev1049_threelevel.mov}{ev1049\_threelevel.mov}: Three-level dynamics showing
a graph moving quickly through assorted configurations.
The coarsest graph is maintained automatically from
modifications the first-level coarsener makes to the
second graph.
\item \demo{ev1049_compare.mov}{ev1049\_compare.mov}: Side-by-side comparison
of single-level vs. three-level dynamics, illustrating
the quicker convergence achieved by multilevel dynamics.
\item \demo{ev1049_multilevel.mov}{ev1049\_multilevel.mov}: Showing quick
convergence times using multilevel dynamics (4-6 levels)
on static graphs being reset to random vertex positions.
\item \demo{ev1049_randomgraph.mov}{ev1049\_randomgraph.mov}: Visualization of the
emergence of the giant component in a random graph
(\refsec{s:randomgraph}).  (8 cores, Euler step).
\item \demo{ev1049_tree.mov}{ev1049\_tree.mov}: Visualization of rapid
insertions into a binary tree (8 cores, Euler step).
\end{itemize}


\section{Related work}

Graph animation is widely used for exploring and
navigating large graphs (e.g., \cite{Herman:IEEETVCG:2000}.)
There is a vast amount of work in this area, so we focus
on systems that aim to animate changing graphs.

Offline graph animation tools develop an animation from a sequence
of key frames (layouts of static graphs).  Such systems find 
layouts for key frame graphs, and then 
interpolate between the key frames in an appropriate way
(e.g. \cite{Friedrich:JGAA:2002}).

In contrast, our system tackles the problem of \emph{online}
or \emph{incremental} graph layout.
We use the term online in the
sense of \emph{online algorithms}:
one receives
a sequence of requests (in our case,
changes to a graph), and each request must be processed
without foreknowledge of future requests.

The key frame approach can
be adapted to address the online problem by computing a new key frame each
time a request arrives, and then interpolating to the new key frame.
For example, Huang et al \cite{Huang:JVLC:1998} developed a system
for browsing large, partially known graphs, where navigation actions
add and remove subgraphs.  They use force-directed
layout for key frame graphs, and interpolate between them.

Another approach is to take an existing graph layout algorithm
and incrementalize (or dynamize) it.  For example, 
the Dynagraph system \cite{North:GD:2001,Ellson:GDS:2004}
uses an incrementalized version of the batch Sugiyama-Tagawa-Toda algorithm
\cite{Sugiyama:TSMC:1981}.

Our basic approach is to develop an incrementalized version
of Walshaw's multilevel force-directed layout algorithm
\cite{Walshaw:JGAA:2003}.  We perform a
continuously running force-directed layout.  When a graph change request
arrives, the graph is immediately updated.  This changes the potential
function, so that the current layout is no longer an equilibrium
position.  The layout immediately starts to seek a new equilibrium.
If the changes come rapidly enough, the graph will be in continuous
motion.  One advantage of this approach is that changes to the graph
are shown immediately, without delaying to compute a key frame.



\section{Layout Dynamics}

\label{s:dynamics}

We use Lagrangian dynamics to derive the equations
of motion for the simulation.
Lagrangian dynamics is a bit excessive for
a simple springs-and-repulsion graph layout.
However, we have found that convergence times
are greatly improved by dynamically adapting
the interpolation between coarser and finer
graphs.  For this we use generalized forces,
which are easily managed with a Lagrangian
approach.

As is conventional, we write $\dot{\bx}_i$
for the velocity of vertex $i$, and
$\ddot{\bx}_i$ for its acceleration.
We take all masses to be 1, so that velocity
and momentum are interchangeable.

In addition to the potential energy $V$ (\refeqn{e:potential}), 
we define a kinetic energy $T$.  For a single graph, this is
simply:
\begin{align}
T &= \sum_{v_i \in V} \tfrac{1}{2} \| \dot{\bx}_i \|^2
\label{e:kinetic}
\end{align}
Roughly speaking, $T$ describes
channels through which potential energy (layout badness)
can be converted to kinetic energy (vertex motion).  
Kinetic energy is then dissipated through friction, which
results in the system settling into a
local minimum of the potential $V$.
We incorporate friction by adding extra
terms to the basic equations of motion.

The equations of motion are obtained
from the Euler-Lagrange equation:
\begin{align}
\left[
\frac{d}{dt} \frac{\partial}{\partial \dot{\bx}_i}
- \frac{\partial}{\partial \bx_i}
\right]
L &= 0
\label{e:eulerlagrange}
\end{align}
where the quantity $L=T-V$ is the Lagrangian.

\subsection{Single level dynamics}

\label{s:singleleveldynamics}
The coarsest graph has straightforward dynamics.
Substituting the definitions of (\ref{e:potential},\ref{e:kinetic})
into the Euler-Lagrange equation yields the basic equation
of motion
$\ddot{\bx}_i = \bF_i$ for a vertex,
where $\bF_i$ is the net force:
\begin{align}
\ddot{\bx}_i &= \underbrace{\sum_{v_i,v_j} - K (\bx_i - \bx_j )}_{\text{spring forces}}
+ \underbrace{\sum_{v_j \neq v_i} \frac{f_0}{(\epsilon_R + \|\bx_i - \bx_j\|)^2 } \cdot \frac{\bx_i - \bx_j}{\| \bx_i - \bx_j \|}}_{\text{repulsion forces}}
\label{e:ddotbx}
\end{align}
We calculate
the pairwise repulsion forces in $O(n \log n)$ time
(with $n=|V|$, the number of vertices)
using the Barnes-Hut algorithm \cite{Barnes:N:1986}.
\ignore{***which has been found to be faster than methods such as Fast Multipole
Method (FMM) when high accuracy is not required
\cite{Blelloch:PA:1997}.
***}

The spring and repulsion forces are supplemented by a damping force
defined by
$\vect{F}^d_i = - d \dot{\vect{x}}_i$
where $d$ is a constant.

Our system optionally adds a `gravity' force that encourages directed
edges to point in a specified direction (e.g., down).


\subsection{Two-level dynamics}
\label{s:twolevel}

We now describe how the dynamics of a graph interacts with its
coarser version.
For notational clarity we write
$\by_i$ for the position of the coarse vertex
corresponding to vertex $i$,
understanding that each vertex in the coarse graph may
correspond to multiple vertices in the finer graph.\footnote{
The alternative to `$\by_i$' would be to write 
e.g. $\bx_{c(i)}^{l+1}$,
meaning the vertex in level $l+1$ corresponding to vertex
$i$ in level $l$.}

In Walshaw's static multilevel layout algorithm \cite{Walshaw:JGAA:2003},
each vertex $\bx_i$ simply uses as its starting position $\by_i$,
the position of its coarser version.
To adapt this idea to a dynamic setting, we
begin by defining the position of $\bx_i$ to be
$\by_i$ plus some displacement $\bdelta_i$, i.e.,:
\begin{align*}
\bx_i &= \delta_i + \by_i
\end{align*}
However, in practice this does not work as well 
as one might hope, and convergence is faster if
one performs
some scaling from the coarse to fine graph,
for example
\begin{align*}
\bx_i &= \delta_i + a \by_i
\end{align*}
A challenge in doing this is that the appropriate
scaling depends on the characteristics of the particular
graph.  Suppose the coarse graph roughly halves the
number of vertices in the fine graph.
If the fine graph is, for example, a three-dimensional
cube arrangement of vertices with 6-neighbours, then the expansion 
ratio needed will be $\approx 2^{1/3}$ or about $1.26$;
a two-dimensional square arrangement of vertices
needs an expansion ratio of $\approx \sqrt{2}$ or about $1.41$.  
Since the graph is dynamic,
the best expansion ratio can also change over time.
Moreover, the optimal amount of scaling might be different for each
axis, and there might be differences in how the fine and coarse
graph are oriented in space.

Such considerations led us to consider affine transformations
from the coarse to fine graph.
We use projections of the form
\begin{align}
\bx_i &= \bdelta_i + \balpha \by_j + \bbeta
\label{e:projection}
\end{align}
where $\balpha$
is a linear transformation (a 3x3 matrix)
and $\bbeta$ is a translation.\ignore{\footnote{By 
gluing together graphs of different dimensionalities it is 
easy to construct examples where no affine
transformation will achieve a good fit.  We believe that
the eventual solution will be to perform some style of
dynamic warping.}}
The variables $(\balpha,\bbeta)$ are themselves given dynamics, 
so that the projection converges to a least-squares fit of
the coarse graph to the fine graph.

\ignore{
One disadvantage of this approach is that it introduces
redundant degrees of freedom (dofs).  For example, the dynamics
are invariant under a transformation 
$\bbeta \mapsto \bbeta + \epsilon$
and $\bdelta_i \mapsto \bdelta_i - \epsilon$
(for each $i$), e.g.,
shifting the projection of the coarse graph to the right
and compensating by moving all the displacements to the left.
Redundant degrees of freedom can cause numerical instability
problems.
In future work we plan to investigate
ways of eliminating the redundant dofs.
}
\subsubsection{Frame dynamics}

\label{s:framedynamics}

We summarize here the derivation of the
time evolution equations for the
affine transformation $(\balpha,\bbeta)$.
Conceptually, we think of the displacements $\delta_i$ as
``pulling'' on the transformation: if all the displacements are to
the right, the transformation will evolve to shift
the coarse graph to the right; if they 
are all outward, the transformation will expand the 
projection of the coarse graph, and so forth.  In this
way the finer graph `pulls' the projection of the
coarse graph around it as tightly as possible.

We derive the equations for $\ddot{\balpha}$ and
$\ddot{\bbeta}$ using Lagrangian dynamics.
To simplify the derivation
we pretend that both graph layouts
are stationary, and that the displacements $\delta_i$
behave like springs between the fine graph and the
projected coarse graph, acting on $\balpha$ and $\bbeta$ 
via `generalized forces.'
By setting up appropriate potential and kinetic
energy terms, the Euler-Lagrange equations yield:
\begin{align}
\ddot{\balpha} &= \tfrac{1}{n} \sum_i \bdelta_i \by_i^T + \by_i \bdelta_i^T \\
\ddot{\bbeta} &= \tfrac{1}{n} \sum_i \bdelta_i
\end{align}
To damp oscillations and dissipate energy 
we introduce damping terms of $- d_\alpha \dot{\balpha}$
and $- d_\beta \dot{\bbeta}$.
\subsubsection{Time dilation}
\label{s:timedilation}

We now turn to the equations of motion for vertices in the
two-level dynamics.  
\ignore{***
A consequence of defining vertex positions
relative to their coarse vertices is that the
the state variables for the two-level dynamics 
are $(\delta_i,\dot{\delta}_i)$ (the relative displacement
and velocity) rather than $(\bx_i,\dot{\bx}_i)$.
***}
The equations for $\dot{\bdelta}_i$ and $\ddot{\bdelta}_i$
are obtained by differentiating \refeqn{e:projection}:
\begin{align}
\bx_i &= \bdelta_i + \underbrace{\bbeta + \balpha \by_i}_{\text{proj. position}} \label{e:bxi} \\
\dot{\bx}_i &= \dot{\bdelta}_i + \underbrace{\dot{\bbeta} + \dot{\balpha} \by_i + \balpha \dot{\by}_i}_\text{proj. velocity} \\
\ddot{\bx}_i &= \ddot{\bdelta}_i + \underbrace{\ddot{\bbeta} + \ddot{\balpha} \by_i + 2 \dot{\balpha} \dot{\by}_i
  + \balpha \ddot{\by}_i}_\text{proj. acceleration}
\label{e:ddotxi}
\end{align}
Let $\bF_i$ be the forces acting on the vertex $\bx_i$.  Substituting
\refeqn{e:ddotxi} into $\ddot{\bx}_i = \bF_i$ and rearranging, one
obtains an equation of motion for the displacement:
\begin{align}
\ddot{\bdelta}_i &= \bF_i - \left( \underbrace{\ddot{\bbeta} + \ddot{\balpha} \by_i + 2 \dot{\balpha} \dot{\by}_i
  + \balpha \ddot{\by}_i}_\text{proj. acceleration} \right)
\label{e:pseudoforce}
\end{align}
The projected acceleration of the coarse vertex
can be interpreted as a `pseudoforce'
causing the vertex to react against motions of its coarser
version.
If \refeqn{e:pseudoforce} were used as the equation
of motion, the coarse and fine graph would evolve independently, with no
interaction.  (We have used this useful property to check correctness of
some aspects of our system.)

\ignore{
A more insidious problem occurs when simulating multiple
levels of coarseness.  (Extra degree of freedom... alternating
signs and blowing up with levels of coarseness.)
}

The challenge, then, is how to adjust \refeqn{e:pseudoforce} in
some meaningful way to couple the finer and coarse graph
dynamics.  Our solution is based on the idea that the
coarser graph layout evolves similarly to the finer graph,
\emph{but on a different time scale}: the coarse graph
generally converges much more quickly.
To achieve a good fit between the coarse and fine graph
we might slow down the evolution of the coarse graph.
Conceptually, we try to do the opposite, speeding up evolution of the fine
graph to achieve a better fit.
\ignore{
}
Rewriting \refeqn{e:bxi} to make each variable an explicit
function of time, and incorporating a time dilation, we obtain
\begin{align}
\bx_i(t) &= \bdelta_i(t) + \underbrace{\bbeta(t) + \balpha(t) \by_i(\phi t)}_{\text{proj. position}} 
\end{align}
where $\phi$ is a time dilation factor to account for the
differing time scales of the coarse and fine graph.  Carrying this through to the
acceleration equation yields the equation of motion
\begin{align}
\ddot{\bdelta}_i &= \bF_i - \left( \underbrace{\ddot{\bbeta} + \ddot{\balpha} \by_i + 2 \dot{\balpha} \phi \dot{\by}_i
  + \balpha \phi^2 \ddot{\by}_i}_\text{proj. acceleration} \right)
\label{e:finalddotdelta}
\end{align}
If for example the coarser graph layout converged at a
rate twice that of the finer graph, we might take
$\phi = \tfrac{1}{2}$, with the effect that we would
discount the projected acceleration $\ddot{\by}_i$ by a factor of
$\phi^2 = \tfrac{1}{4}$.  
In practice we have used values of $0.1 \leq \phi \leq 0.25$.
Applied across multiple levels of coarse graphs,
we call this approach \emph{multilevel time dilation}.

\ignore{
One concern in choosing $\phi$ is that it be
small enough that errors do not grow exponentially
across levels.  Consider for example a small error
$\epsilon$ in $\ddot{y_i}$, so that the 
last term of \refeqn{e:finalddotdelta} is
$\balpha \phi^2 (\ddot{\by}_i + \epsilon)$.
The magnitude of error
propagated to $\ddot{\delta}_i$ can be
$\lambda_1 \phi^2 \| \epsilon \|$, where $\lambda_1$ is the
principal eigenvalue of $\alpha$.
If $\phi > \frac{1}{\sqrt{\lambda_k}}$ then
the error magnitude will increase.
Generally we expect that the coarser graph will be
expanded by at most a factor of two,
so that $\lambda_1 \leq 2$.  By this reckoning
choosing $\phi < \frac{1}{\sqrt{2}}$ is
reasonable.
In practice we have used $\phi = 0.1$ and
$\phi = 0$; both yield reasonable results.
More experimentation is needed.
}

In addition to the spring and repulsion forces in
$\bF_i$, we include a drag term 
$\bF_i^d = -d \dot{\bdelta}_i$
in the forces of \refeqn{e:finalddotdelta}.

\subsection{Multilevel dynamics}
To handle multiple levels of coarse graphs, we iterate the
two-level dynamics.
The dynamics simulation simultaneously integrates the following
equations:
\begin{itemize}
\item The equations of motion for the vertices in the coarsest
graph, using the single-level dynamics of \refsec{s:singleleveldynamics}.
\item The equations for the projection $\balpha, \bbeta$ between each
coarser and finer graph pair (\refsec{s:framedynamics}).
\item The equations of motion $\ddot{\bdelta}_i$ for the
displacements of vertices in the finer graphs, using the
two-level dynamics of \refsec{s:twolevel}.
\end{itemize}

In our implementation, the equations are integrated using an explicit,
fourth-order Runge-Kutta method.  (We also have a simple Euler-step
method, which is fast, but not as reliably stable.)


\subsection{Equilibrium positions of the multilevel dynamics}
\label{s:equilibria}

We prove here that a layout found using the
multilevel dynamics is an equilibrium position
of the potential energy function of \refeqn{e:potential}.
This establishes that the multilevel approach
does not introduce spurious minima, and can 
be expected to converge to the same layouts
as a single-level layout, only faster.

\vspace{1em}
\begin{thm}
\label{thm:equilibria}
Let $(X,\dot{X})$ be an equilibrium position of the two-level
dynamics, where $X = (\bdelta_1, \bdelta_2, \ldots,
\balpha, \bbeta, \by_1, \by_2, \ldots)$,
and $\ddot{X} = \dot{X} = 0$.
Then, $(\bx_1,\bx_2,\ldots,\bx_n)$ is an equilibrium position
of the single-level dynamics, 
where $\bx_i = \bdelta_i + \balpha \by_i + \bbeta$,
and the single-level potential gradient $\nabla V = 0$ 
(\refeqn{e:potential}) vanishes there.
\end{thm}

\begin{proof}
Since $\dot{X}=0$, the drag terms vanish
from all equations of motion.
Substituting $\dot{X}=0$ and $\ddot{X}=0$ into
\refeqn{e:finalddotdelta} yields
$\bF_i = 0$ for each vertex.  
Now consider the single-level
dynamics (\refsec{s:singleleveldynamics})
using the positions
$\bx_i$
obtained from 
$\bx_i = \delta_i + \balpha y_i + \bbeta$ (\refeqn{e:projection}).
From $\ddot{\bx}_i = \bF_i$
we have
have $\ddot{\bx}_i = 0$ for each $i$.
The Euler-Lagrange equations for the
single level layout are
(\refeqn{e:eulerlagrange}):
\begin{align*}
\frac{d}{dt} \frac{\partial L}{\partial \dot{\bx_i}} 
- \frac{\partial}{\partial \bx_i} L
&= 0
\end{align*}
Since $\frac{d}{dt} \frac{\partial L}{\partial \dot{\bx_i}} = \ddot{\bx_i} = 0$,
we have
$-\frac{\partial}{\partial \bx_i} L = 0$.
Using $L=T-V$ and that the kinetic energy $T$ does not
depend on $\bx_i$, we obtain
\begin{align*}
\frac{\partial}{\partial \bx_i} V = 0
\end{align*}
for each $i$.  Therefore $\nabla V = 0$ at this point.
\end{proof}
\ignore{
(Can we establish that it isn't a maximum or a saddle point?
We would need to use the fact that $X$ is a stable equilibrium
and make a comment about the positive definiteness...?)
}

This result can be applied inductively over pairs
of finer-coarser graphs, so that \refthm{thm:equilibria} holds also for
multilevel dynamics.


\section{Dynamic coarsening}


\label{s:coarsening}

As vertices and edges are added to and removed from
the base graph
$G=(V,E)$,
our system dynamically maintains the coarser graphs
$G_1,G_2,\ldots,G_m$.
Each vertex in a coarse graph may correspond to
several vertices in the base graph, which is to say,
each coarse graph defines a partition of the
vertices in the base graph.  It is useful
to describe coarse vertices as subsets of $V$.
For convenience we define a finest graph $G_0$
isomorphic to $G$, with vertices $V_0 = \{ \{v\} ~:~ v \in V \}$,
and edges $E_0 = \{ (\{v_1\},\{v_2\}) ~:~ (v_1,v_2) \in E\}$.
Coarser graphs are obtained by merging vertices via
set union.  The sequence $V_0,V_1,\ldots,V_m$ of
of coarse graph vertices forms a chain in the
lattice of partitions of $V$; we need to maintain
this chain of partitions as changes are made to the base graph
$V_0$.

We have devised an algorithm
that efficiently maintains $G_{i+1}$ in response to changes
in $G_i$.  By applying this algorithm at each level
the entire chain $G_1,G_2,\ldots,G_m$ is
maintained.

We present a fully dynamic, Las Vegas-style randomized graph
algorithm for maintaining a coarsened version of a graph.  For
graphs of bounded degree, this algorithm requires $O(1)$ operations
on average per edge insertion or removal.

Our algorithm is based on the traditional
matching approach to coarsening developed by Hendrickson and Leland
\cite{Hendrickson:SIAMSC:1995}.  
Recall that a matching of a graph $G=(V,E)$ is a
subset of edges $M \subseteq E$ satisfying the restriction that no
two edges of $M$ share a common vertex.  A matching is \emph{maximal}
if it is not properly contained in another matching.
A maximal matching can be found by considering
edges one at a time and adding them if they do not conflict
with an edge already in $M$.
(The problem of finding a maximal matching 
should not be confused with that of finding a 
\emph{maximum cardinality} matching, 
a more difficult problem.)

\ignore{***
\begin{figure}
\begin{align*}
\newcommand{\node}{*+[o][F-]{ }}
\newcommand{\edge}[1]{\ar@{-}[#1]}
\newcommand{\boldedge}[1]{\ar@{=}[#1]}
\xymatrix @=0.5cm {
    & \{ v_1 \} \boldedge{d}      & \\
\{ v_5 \} \edge{ur}1 & \{ v_2 \} \edge{d} \edge{l} & \{v_5\} \boldedge{r} & \{v_1,v_2\} \edge{d} & \{v_1,v_2,v_5\} \edge{d} \\
    & \{ v_3 \} \boldedge{d}      &                  & \{v_3,v_4\}  & \{v_3,v_4\} \\
    & \{ v_4 \} \\
& (a) & & (b) & (c)
}
\end{align*}
\caption{Illustration of multilevel coarsening by matching.
Double lines indicate matched edges.
(a) the initial graph $V_0$ obtained from the original
graph by replacing each vertex $v_i$ with the singleton
set $\{v_i\}$; (b) after contracting the matched edges
of (a); (c) after contracting the matched edges of
(b).
}
\end{figure}
***}

\subsection{Dynamically maintaining the matching}

We begin by making
the matching unique.
We do this by fixing a total order $<$ on the
edges, chosen uniformly at random.
(In practice, we compute $<$ using a bijective hash function.)
To produce a matching we can consider each 
edge in ascending order by $<$,
adding it to the
matching if it does not conflict with a previously matched edge.
If $e_1 < e_2$, we say that $e_1$ has priority over $e_2$ for matching.

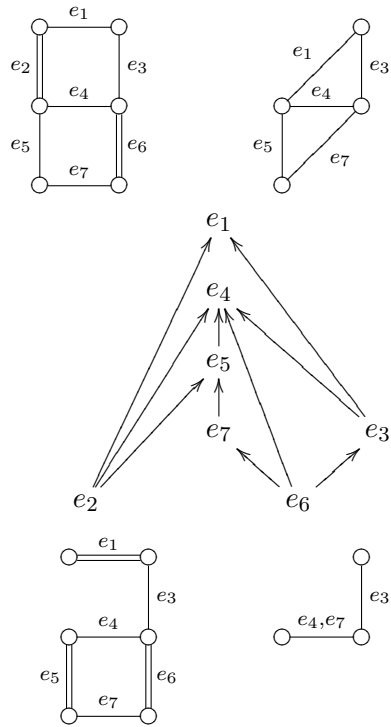
\begin{figure}
\newcommand{\node}{*+[o][F-]{ }} 
\newcommand{\edge}[1]{\ar@{-}[#1]}
\newcommand{\boldedge}[1]{\ar@{=}[#1]}
\begin{align*}
\xymatrix{
\node \edge{r}^{e_1} \boldedge{d}_{e_2} & \node \edge{d}^{e_3} \\
\node \edge{r}^{e_4} \edge{d}_{e_5} & \node \boldedge{d}^{e_6} \\
\node \edge{r}^{e_7} & \node
}~~~~~~~~~~~
\xymatrix{
& \node \edge{dl}_{e_1} \edge{d}^{e_3} \\ 
\node \edge{d}_{e_5} \edge{r}^{e_4} & \node \edge{dl}^{e_7} \\
\node
} \\
\xymatrix @=0.5cm {
& & e_1 \\
& & e_4 \\
& & e_5 \ar[u] \\
& & e_7 \ar[u] & & e_3 \ar[uull] \ar[uuull] \\
e_2 \ar[uurr] \ar[uuurr] \ar[uuuurr] & & & e_6 \ar[ul] \ar[uuul] \ar[ur]
} \\
\xymatrix{
\node \boldedge{r}^{e_1} & \node \edge{d}^{e_3} \\
\node \edge{r}^{e_4} \boldedge{d}_{e_5} & \node \boldedge{d}^{e_6} \\
\node \edge{r}^{e_7} & \node
}~~~~~~~~~~~
\xymatrix{
& \node \edge{d}^{e_3} \\
\node \edge{r}^{e_4,e_7} & \node
} \\
\end{align*}
\caption{\label{f:dualgraph}Illustration of dynamic
coarsening.  Top left: A graph with double lines
indicating matched edges;
Top right: its coarsened version, obtained by contracting matched
edges; Middle: the edge dependence graph for the initial graph,
using the (arbitrarily chosen)
order $e_6 < e_2 < e_3 < e_7 < e_5 < e_1 < e_4$.
Initially the matching is $\{e_2,e_6\}$.
Removing edge $e_2$ (bottom left)
triggers re-evaluation of the
match equations for $e_5$, $e_1$, and $e_4$.
The new matching is $\{e_6,e_5,e_1\}$,
resulting in changes to the coarsened graph (bottom right).
}
\end{figure}

Our basic analysis tool is the \emph{edge graph}
$G^\ast = (E,S)$
whose vertices are the edges of $G$, and
$e_1 S e_2$ when the edges share a vertex.
A set of edges $M$ is a matching on $G$ if and only
if $M$is an independent set of vertices in $G^\ast$.
From $G^\ast$ we can define an
\emph{edge dependence graph}
$\mathcal{E} = (E,\rightarrow)$
which is a directed version of $G^\ast$:
\begin{align*}
e_1 \rightarrow e_2 &\equiv (e_1 < e_2) 
~\text{and}~e_1 S e_2~\text{(share a common vertex)}
\end{align*}
Since $<$ is a total order, the edge dependence graph
$\mathcal{E}$ is acyclic.  \reffig{f:dualgraph}
shows an example.

Building a matching by considering the edges in order of $<$ is
equivalent to a simple rule: $e$ is matched if and only if there
is no edge $e' \in M$ such that $e' \rightarrow e$.
We can express this rule as a set of \emph{match equations} 
whose solution can be
maintained by simple change propagation.
Let $m : E \rightarrow \{\bot,\top\}$ indicate whether
an edge is matched: $m(e) = \top$ if and only if $e \in M$.
The match equation for an edge $e \in E$ is:
\begin{align*}
m(e) &= \bigwedge_{e' ~:~ e' \rightarrow e} \neg m(e')
\end{align*}
where by convention $\bigwedge \emptyset = \top$.

To evaluate the match equations we place
the edges to be considered for matching in a priority queue 
ordered by $<$, so that highest priority
edges are considered first.
The match equations can then be evaluated using a 
straightforward change propagation algorithm:

\noindent
While the priority queue is not empty:
\begin{enumerate}
\item Retrieve the highest priority edge $e=(v_1,v_2)$ from the
queue and evaluate its match equation $m(e)$.
\item If the solution of the match equation has changed, then:
\begin{enumerate}
\item If $m(e) = \top$, then $\mathrm{match}(v_1,v_2)$.
\item If $m(e) = \bot$, then $\mathrm{unmatch}(v_1,v_2)$.
\end{enumerate}
\end{enumerate}
Both $\mathrm{match}(e)$ and $\mathrm{unmatch}(e)$
add the dependent edges of $e$ to the queue, so
that changes ripple through the graph.

\ignore{Maybe start with a formal statement of invariants,
and derive each of these carefully from that statement.}

\reffig{f:matchops}
summarizes the basic steps required to
maintain the coarser graph ($V',E'$) as edges and vertices are added
and removed to the finer graph.  

\begin{figure}
\hrule\hfill

\vspace{1em}
\begin{centering}
{\sc Dynamic-Match} operations

\end{centering}

We use $\overline{v}$ to indicate the coarse
version of a vertex $v$.

\begin{itemize}
\item addEdge(e): Increase the count
of $(\overline{v_1},\overline{v_2})$ in $E'$ (possibly adding edge if not
already present).  Add $e$ to the queue.
\item deleteEdge(e):
If $e$ is in the matching then $\mathrm{unmatch}(e)$.
Decrease the count of $(\overline{v_1},\overline{v_2})$ in $E'$; if this 
count is 0 then delete this edge from $E'$.
If either $v_1$ or $v_2$ have no edges except $e$, then
remove $\overline{v_1}$ (resp. $\overline{v_2}$) from the coarse
graph.
Add all edges $e'$ with $e \rightarrow e'$ to the queue.
\item deleteVertex(v): for each edge $e$ incident
on $v$, $\mathrm{deleteEdge}(e)$.
Remove $v$ from $G$.  
\item addVertex(v): no action needed.
\item $\mathrm{match}(e)$,
where $e = (v_1,v_2)$:
For each edge $e'$ where $e \rightarrow e'$, if
$e'$ is matched then $\mathrm{unmatch}(e')$.
Delete vertices $\overline{v_1}$ and $\overline{v_2}$ from the
coarser graph.
Create a new vertex $v_1 \cup v_2$ in $G'$.
For all edges $e=(v,v')$ in G incident on $v_1$ or $v_2$
(but not both), add a corresponding edge to or from
$v_1 \cup v_2$ in $G'$.
For each $e'$ such that $e \rightarrow e'$,
add $e'$ to the queue.
\item $\mathrm{unmatch}(e)$,
where $e = (v_1,v_2)$:
delete any edges in $G'$ incident on $v_1 \cup v_2$.
Delete the vertex $v_1 \cup v_2$ from $G'$.
Add new vertices $\overline{v_1}$ and $\overline{v_2}$ to
$G'$.  For each edge incident on $v_1$ or $v_2$
in $G$, add a corresponding edge to $G'$.
For each $e'$ such that $e \rightarrow e'$,
add $e'$ to the queue.

\vspace{1em}
\hrule\hfill
\end{itemize}
\caption{\label{f:matchops}Basic operations of the dynamic
coarsening algorithm.}
\end{figure}


\subsection{Complexity of the dynamic matching}

The following theorem establishes that
for graphs of bounded degree,
the expected cost of dynamically
maintaining the coarsened graph is $O(1)$ per edge inserted or removed
in the fine graph.
The cost does not depend on the number of edges in the graph.

\vspace{1em}
\begin{thm}
\label{thm:matchcomplexity}
In a graph of maximum degree $d$, 
the randomized complexity of $\mathrm{addEdge}(e)$ and $\mathrm{removeEdge}(e)$
is $O(d e^{2d})$.
\end{thm}

We first prove a lemma concerning the extent to which
updates may need to propagate in the
edge dependence graph.  As usual for randomized algorithms, we analyze 
the complexity using ``worst-case average time,'' i.e., the maximum
(with respect to choices of edge to add or remove) of the expected time 
(with expectation taken over random priority assignments).
For reasons that will become clear we define the priority order $<$
by assigning to each edge 
a random real in $[0,1]$, with $1$ being the highest priority.

\vspace{1em}
\begin{lem}
\label{lem:matchcomplexity}
Let $G^\ast=(E,S)$ be a graph 
and $\rho : E \rightarrow [0,1]$ an injective function
chosen uniformly at random
assigning to each vertex a real priority in $[0,1]$.
If $G^\ast$ has maximum degree $k$, the
expected number of vertices reachable from any vertex  in $E$
following only edges from higher to lower priority vertices
is $O(e^k)$.
\end{lem}

\begin{figure}
\vspace{0.3in}
\begin{align*}
\newcommand{\node}{*+[o][F-]{ }}
\xymatrix @=0.5cm {
e_4 & e_1 & e_5 \ar@/^1pc/[ll] & e_7 \ar[l] & e_3 \ar@/^2pc/[lll] \ar@/^1pc/[ll] & e_2 \ar@/_1pc/[lll] \ar@/_3pc/[lllll] \ar@/_2pc/[llll] & e_6 \ar[l] \ar@/^1pc/[ll] \ar@/^3pc/[llllll] 
}
\end{align*}
\vspace{0.3in}
\caption{\label{f:lineararrangement}A linear arrangement
of the edge dependence graph of \reffig{f:dualgraph}.}
\end{figure}

\begin{proof}
It is helpful to view the priority assignment
$\rho$ as inducing a linear arrangement of the
vertices, i.e., we might draw $G^\ast$
by placing its vertices on the real line at their priorities.
We obtain a directed graph $(E,\rightarrow)$ by making edges always point
toward zero, i.e., from higher to lower priorities
(cf. \reffig{f:lineararrangement}).
Note that vertices with low priorities will tend to have
high indegree and low outdegree.

\vspace{1em}
We write $\mathbb{E}[\cdot]$ for expectation with
respect to the random priorities $\rho$.
For $e \in E$, let $N(e)$ be the number of 
vertices of $G^\ast$ reachable from $e$ by following
paths that move from higher to lower priority vertices.
We bound the expected value of $N(e)$ given
its priority $\rho(e) = \eta$: we can always
reach $e$ from itself, and we can follow any
edges to lower priorities:
\begin{align}
\mathbb{E}[N(e) ~|~ \rho(e) = \eta] 
&\leq 
1 + \sum_{e' ~:~ e S e'} 
\underbrace{\mathrm{Pr}(\rho(e') < \eta)}_{= \eta} 
\cdot \mathbb{E}[N(e') ~|~ \rho(e') < \eta]
\label{e:matchcomp1}
\end{align}
Let $f(\eta) = \sup_{e \in E} \mathbb{E}[N(e) ~|~ \rho(e) = \eta]$.
Then, 
\begin{align}
\mathbb{E}[N(e') ~|~ \rho(e') < \eta]
&\leq \int_{0}^\eta \eta^{-1} f(\alpha) d\alpha
\label{e:matchcomp2}
\end{align}
where the integration averages $f$ over a uniform
distribution on priorities $[0,\eta)$.
Since the degree of any vertex is $\leq k$,
there can be at most $k$ terms in the
summation of \refeqn{e:matchcomp1}.
Combining the above, we obtain
\begin{align}
f(\eta) &= \sup_{e \in E} \mathbb{E}[N(e) ~|~ \rho(e) = \eta] \\
        &\leq \sup_{e \in E} \left(1 + \sum_{e' S e} \eta \mathbb{E}[N(e') ~|~ \rho(e') < \eta]\right) \\
        &\leq 1 + k \eta \int_0^\eta \eta^{-1} f(\alpha) d\alpha \\
        &\leq 1 + k \int_0^\eta f(\alpha) d\alpha
\label{e:matchcomp3}
\end{align}
Therefore $f(\eta) \leq g(\eta)$, where $g$ is
the solution to the integral equation
\begin{align}
g(\eta) &= 1 + k \int_0^\eta g(\alpha) d\alpha
\label{e:xxxx}
\end{align}
Isolating the integral and differentiating yields
the ODE $g(\eta) = k^{-1} g'(\eta)$, which
has the solution $g(\eta) = e^{\eta k}$,
using the boundary condition $g(0)=1$
obtained from \refeqn{e:xxxx}.
Since $0 \leq \eta \leq 1$,
$g(\eta) \leq e^k$.
Therefore, for every $e \in E$,
the number of reachable vertices
satisfies $\mathbb{E}[N(e)] \leq e^k$.
\end{proof}

Note that the upper bound of $O(e^k)$ vertices reachable
depends only on the maximum degree, and not on the size
of the graph.

We now prove \refthm{thm:matchcomplexity}.

\begin{proof}
If a graph $G=(V,E)$ has maximum degree $d$, its edge
graph $G^\ast$ has maximum degree $2(d-1)$.  Inserting
or removing an edge
will cause us to reconsider the matching of $\leq e^{2(d-1)}$
edges on averages by \reflem{lem:matchcomplexity}.
If a max heap is used to implement the priority queue,
$O(d e^{2d})$ operations are needed to insert and
remove these edges.
Therefore the randomized complexity is $O(d e^{2d})$.
\end{proof}

In future work we hope to extend our analysis
to show that the entire sequence of coarse graphs
$G_1,G_2,\ldots,G_m$ can be efficiently maintained.
In practice, iterating the algorithm described here
appears to work very well.


\section{Implementation}

Our system is implemented in C++, using OpenGL and pthreads.
The graph animator runs in a separate thread from the user
threads.  The basic API is simple, with methods
$\mathrm{newVertex}()$ and $\mathrm{newEdge}(v_1,v_2)$
handling vertex and edge creation, and destructors
handling their removal.  
\ignore{
\begin{figure}
\begin{verbatim}
layout_t myGraph;

int main(int argc, char** argv)
{
  pthread_t thread;
  pthread_create(&thread, 0, 
    userThread, 0);

  GraphAnimator<layout_t>* ga = new 
    GraphAnimator<layout_t>(argc, argv, myGraph);
  ga->mainLoop();
}

void* userThread(void*)
{
  // Make a ring of vertices.
  const int N = 10;
  vertex_t* v[N];
  for (int i=0; i < N; ++i)
    v[i] = myGraph.newVertex();

  // Add the edges
  for (int j=0; j < N; ++j)
  {
    int nj = (j+1) % N;
    edge_t* e = myGraph.newEdge(v[j], v[nj]);
    sleep(1);
  }
}
\end{verbatim}
\caption{\label{f:hellograph}A simple program that
creates a ring graph with 20 vertices.  Include 
statements are omitted.
}
\end{figure}
}

For static graphs, 
we have so far successfully used up to six levels of
coarsening, with the coarsened graphs computed in
advance.  With more than six levels we are encountering
numerical stability problems that seem to be related
to the projection dynamics. 

For dynamic graphs we have used three levels (the
base graph plus two coarser versions), with the
third-level graph being maintained from the actions
of the dynamic coarsener for the first-level graph.
At four levels we encounter a subtle bug in our
dynamic coarsening implementation we have not yet resolved.


\subsection{Parallelization}
\ignore{***
We currently perform one time step per frame.  In the future
we plan to let the rendering and time stepping proceed at their
own pace, without synchronization, to achieve steady
frame rates.
***}

Our single-level dynamics implementation is parallelized.
Each frame entails two expensive operations: rendering and force
calculations.  We use the Barnes-Hut tree to divide the force
calculations evenly among the worker threads; this results in good
locality of reference, since vertices that interact through
edge forces or near-field repulsions are often handled by
the same thread.
Rendering is performed in a separate thread, with time
step $t$ being rendered while step $t+\delta t$ is
being computed.  The accompanying animations were
rendered on an 8-core (2x4) iMac using OpenGL, compiled
with g++ at -O3.

Our multilevel dynamics engine is not yet parallelized,
so the accompanying demonstrations of this are rendered on 
a single core.  Parallelizing the multilevel dynamics
engine remains for future work.


\section{Applications}

\label{s:randomgraph}

We include with this paper two demonstrations of applications:
\begin{itemize}
\item The emergence of the giant component in a random graph:
In Erd{\"o}s-Renyi $G(n,p)$ random graphs on $n$ vertices
where each edge is present independently with probability $p$, 
there are a number of interesting phase transitions:
when $p < n^{-1}$ the largest connected component is almost
surely of size $\Theta(\log n)$; when $p=n^{-1}$ it is a.s. of
size $\Theta(n^{2/3})$, and when $p > n^{-1}$ it is a.s. of
size $\Theta(n)$---the ``giant component.''
In this demonstration a large random graph is constructed
by preassigning to all ${n \choose 2}$ edges a probability
trigger in $[0,1]$, and then slowly raising a probability parameter
$p(t)$ from $0$ to $1$ as the simulation progresses, with
edges `turning on' when their trigger is exceeded.
\item Visualization of insertions of random elements into
a binary tree, with an increasingly rapid rate of insertions.
\end{itemize}

In addition, we mention that the graph visualizer was of
great use in debugging itself, particularly in
tracking down errors in the dynamic matching implementation.

\section{Conclusions}

We have described a novel approach to real-time visualization of dynamic
graphs.  Our approach combines the benefits of multilevel force-directed
graph layout with the ability to render rapidly changing graphs in
real time.  We have also contributed a novel and efficient method
for dynamically maintaining coarser versions of a graph.


\newcommand{\etalchar}[1]{$^{#1}$}


\end{document}
